\documentclass[namedreferences]{solarphysics}
\usepackage[hyperref,optionalrh,solaromanenum]{spr-sola-addons_arxiv}  
\usepackage{graphicx} 
\usepackage{color}
\usepackage{breakurl}
\usepackage{acronym} 

\hypersetup{
  colorlinks=true,   
  urlcolor=blue,     
  linkcolor=blue,    
  citecolor=blue,    
}

\usepackage{twoopt}  

\definecolor{bllsht}{rgb}{0.40, 0.05, 0.01} 
\long\def\TODO#1TODO{\par\vspace{0.5ex}\noindent
  {\Large \textcolor{red}{@~}}\color{bllsht}{#1}\color{black}\\[0.5ex]}

\definecolor{rrblue}{rgb}{0.15,0.0,0.8} 

\makeatletter
\newcommand{\bibnote}[2]{\global\@namedef{#1note}{#2~}}
\newcommand{\biblink}[2]{\global\@namedef{#1link}{#2}}
\makeatother

\definecolor{amber}{rgb}{1.0, 0.49, 0.0}

\bibpunct{(}{)}{;}{a}{}{,}    
\makeatletter
  \protected\def\sppresslink{\def\hyper@linkstart##1##2{}\let\hyper@linkend\@empty}
  \newcommandtwoopt{\citeads}[3][][]{%
   \href{http://ui.adsabs.harvard.edu/abs/#3/abstract}%
        {\sppresslink \citealp[#1][#2]{#3}}
   \biblink{#3}{\href{http://ui.adsabs.harvard.edu/abs/#3/abstract}{ADS}}}
 \newcommandtwoopt{\citepads}[3][][]{%
   \href{http://ui.adsabs.harvard.edu/abs/#3/abstract}%
        {\sppresslink \citep[#1][#2]{#3}}
   \biblink{#3}{\href{http://ui.adsabs.harvard.edu/abs/#3/abstract}{ADS}}}
 \newcommandtwoopt{\citetads}[3][][]{%
   \href{http://ui.adsabs.harvard.edu/abs/#3/abstract}%
        {\sppresslink \citet[#1][#2]{#3}}
  \biblink{#3}{\href{http://ui.adsabs.harvard.edu/abs/#3/abstract}{ADS}}}
 \newcommandtwoopt{\citeyearads}[3][][]{%
   \href{http://ui.adsabs.harvard.edu/abs/#3/abstract}%
        {\sppresslink \citeyear[#1][#2]{#3}}
   \biblink{#3}{\href{http://ui.adsabs.harvard.edu/abs/#3/abstract}{ADS}}}
\makeatother

\newcount\longrefs
\def\adv{\ifnum\longrefs=1 {Adv.\ Space Res.} \else 
                           {Adv.\ Sp'\ Res.}\fi}
\def\aap{\ifnum\longrefs=1 {Astron.\ Astrophys.}\else 
                           {A\hbox{\rm \&}A}\fi}
\def\aapr{\ifnum\longrefs=1 {Astron.\ Astrophys.\ Rev.}\else 
                            {A\hbox{\rm \&}AR}\fi}
\def\aaps{\ifnum\longrefs=1 {Astron.\ Astrophys.\ Suppl.}\else 
                            {A\hbox{\rm \&}A Suppl.}\fi}
\def\actaa{\ifnum\longrefs=1 {Acta Astronomica}\else
                            {Acta Astron.}\fi}
\def\aipcs{\ifnum\longrefs=1 {Am.\ Inst.\ Phys.\ Conf.\ Series}\else
                             {AIP Conf.\ Ser.}\fi}
\def\aj{\ifnum\longrefs=1 {Astron.\ J.}\else 
                          {AJ}\fi} 
\def\ao{\ifnum\longrefs=1 {Applied Optics}\else 
                           {Appl.\ Opt.}\fi} 
\def\aspcs{\ifnum\longrefs=1 {Astron.\ Soc.\ Pacific Conf.\ Series}\else 
                           {ASP Conf.\ Ser.}\fi} 
\def\apj{\ifnum\longrefs=1 {Astrophys.\ J.}\else 
                           {ApJ}\fi} 
\def\apjl{\ifnum\longrefs=1 {Astrophys.\ J. Lett.}\else 
                            {ApJL}\fi} 
\def\aplett{\ifnum\longrefs=1 {Astrophys.\ J. Lett.}\else 
                            {ApJ}\fi} 
\def\apjs{\ifnum\longrefs=1 {Astrophys.\ J. Suppl.}\else 
                            {ApJS}\fi}
\def\apss{\ifnum\longrefs=1 {Astrophys.\ Space Sci.}\else 
                            {Astrophys.\ Space Sci.}\fi}
\def\araa{\ifnum\longrefs=1 {Ann.\ Rev.\ Astron.\ Astrophys.}\else 
                            {ARA\hbox{\rm \&}A}\fi}
\def\azh{\ifnum\longrefs=1 {Astronomicheskii Zhurnal}\else 
                            {Astron.\ Zhur.}\fi}
\def\baas{\ifnum\longrefs=1 {Bull.\ Am.\ Astron.\ Soc.}\else 
                            {BAAS}\fi}
\def\bain{\ifnum\longrefs=1 {Bull.\ Astronom.\ Institutes Netherlands}\else
                            {Bull.\ Astr.\ Inst.\ Neth.}\fi}
\def\cjaa{\ifnum\longrefs=1 {Chin.\ J.\ Astron.\ Astrophys.}\else 
                            {Chin.\ J.\ A\&A}\fi}
\def\gca{\ifnum\longrefs=1 {Geochim.\ Cosmochim.\ Acta}\else 
                           {Geochim.\ Cosmochim.\ Acta}\fi}
\def\grl{\ifnum\longrefs=1 {Geophys.\ Res.\ Lett.}\else 
                           {Geoph.\ Res.\ Lett.}\fi}
\def\iaucirc{\ifnum\longrefs=1 {IAU Circulars}\else 
                          {IAU Circ.}\fi}
\def\icarus{\ifnum\longrefs=1 {Icarus}\else 
                          {Icarus}\fi}
\def\ip{\ifnum\longrefs=1 {in press}\else 
                          {in press}\fi}
\def\jcap{\ifnum\longrefs=1 {Jour.\ Cosmology Astropart.\ Phys.}\else 
                          {JCAP}\fi}
\def\jgr{\ifnum\longrefs=1 {J.\ Geophys.\ Res.}\else 
                           {J.\ Geophys.\ Res.}\fi}  
\def\jrasc{\ifnum\longrefs=1 {J.\ Royal Astron.\ Soc.\ Canada}\else 
                             {JRAS Can.}\fi}  
\def\memsai{\ifnum\longrefs=1 {Mem.~Soc.~Astron.~Italiana}\else
                              {MmSAI}\fi}
\def\mnras{\ifnum\longrefs=1 {Mon.\ Not.\ Roy.\ Astron.\ Soc.}\else 
                             {MNRAS}\fi} 
\def\na{\ifnum\longrefs=1 {New Astronomy}\else 
                          {New Astron.}\fi}
\def\nar{\ifnum\longrefs=1 {New Astronomy rev.}\else 
                           {New Astron.\ Rev.}\fi}
\def\nat{\ifnum\longrefs=1 {Nature}\else 
                           {Nat}\fi}
\def\pasa{\ifnum\longrefs=1 {Pub.\ Astron.\ Soc.\ Australia}\else 
                            {PASA}\fi} 
\def\pasj{\ifnum\longrefs=1 {Pub.\ Astron.\ Soc.\ Japan}\else 
                            {PASJ}\fi} 
\def\pasp{\ifnum\longrefs=1 {Pub.\ Astron.\ Soc.\ Pacific}\else 
                            {PASP}\fi} 
\def\physscr{\ifnum\longrefs=1 {Physica Scripta}\else 
                               {Phys.\ Scrip.}\fi} 
\def\planss{\ifnum\longrefs=1 {Planetary \& Space Science}\else 
                              {Plan. \& Space Sci.}\fi} 
\def\pre{\ifnum\longrefs=1 {Phys.\ Rev.\ E}\else
                           {Phys.\ Rev.\ E}\fi}
\def\procspie{\ifnum\longrefs=1 {Proc.\ SPIE}\else 
                                {Proc.\ SPIE}\fi} 
\def\qjras{\ifnum\longrefs=1 {Quarterly J.\ Royal Astron.\ Soc.}\else 
                             {QJRAS}\fi} 
\def\rmxaa{\ifnum\longrefs=1 {Revista Mexicana de Astron.\ y Astrofys.}\else 
                             {RMxAA}\fi} 
\def\sa{\ifnum\longrefs=1 {Soviet Astron..}\else 
                          {Sov.\ Astron.}\fi}
\def\skytel{\ifnum\longrefs=1 {Sky \& Telescope}\else 
                              {Sky \& Tel.}\fi} 
\def\solphys{\ifnum\longrefs=1 {Solar Phys.}\else 
                               {SoPh}\fi}
\def\sovast{\ifnum\longrefs=1 {Soviet Astron.}\else 
                              {Sov.\ Ast.}\fi}
\def\ssr{\ifnum\longrefs=1 {Space Sci.\ Rev.}\else 
                           {Space Sci.\ Rev.}\fi}
\def\zap{\ifnum\longrefs=1 {Zeit.\ f.\ Astrophys.}\else
                               {Z.\ Astrophys.}\fi}

\newacro{AA}{Astronomy \& Astrophysics}  
\newacro{ADS}{Astrophysics Data System}
\newacro{AIA}{Atmospheric Imaging Assembly}
\newacro{ALMA}{Atacama Large Millimeter/submillimeter Array}
\newacro{AO}{adaptive optics}
\newacro{ApJ}{Astrophysical Journal}
\newacro{AR}{active region}
\newacro{bb}{bound-bound}
\newacro{bf}{bound-free}
\newacro{BFI}{Broad-band Filter Imager}
\newacro{CE}{coronal equilibrium}
\newacro{CfA}{Center for Astrophysics}
\newacro{CME}{coronal mass ejection}
\newacro{CRD}{complete redistribution}
\newacro{CRISP}{CRisp Imaging SpectroPolarimeter}
\newacro{CRISPEX}{CRisp SPectral EXplorer}
\newacro{CS}{coherent scattering}
\newacro{DEM}{Differential Emission Measure}
\newacro{DF}{dynamic fibril}
\newacro{DKIST}{Daniel K. Inouye Solar Telescope}
\newacro{DLR}{Deutsches Zentrum f\"ur Luft- und Raumfahrt}
\newacro{DOT}{Dutch Open Telescope}
\newacro{DST}{Richard B. Dunn Solar Telescope}   
\newacro{EB}{Ellerman bomb}
\newacro{EDP}{\'{E}dition Diffusion Presse}  
\newacro{EIT}{Extreme ultraviolet Imaging Telescope}
\newacro{EPIC}{European participation in Solar-C}
\newacro{ERC}{European Research Council}
\newacro{ESA}{European Space Agency}
\newacro{EST}{European Solar Telescope}
\newacro{EUV}{extreme ultraviolet}
\newacro{FAF}{flaring active-region fibril}
\newacro{ff}{free-free}
\newacro{FITS}{Flexible Image Transport System}
\newacro{FOV}{field of view}
\newacro{fov}{field of view}
\newacro{FWHM}{full width at half maximum}
\newacro{HAO}{High Altitude Observatory}
\newacro{HD}{hydrodynamics}
\newacro{Hi-C}{High Resolution Coronal Imager Sounding Rocket}
\newacro{HMI}{Helioseismic and Magnetic Imager}
\newacro{IAA}{Instituto de Astrof\'{i}sica de Andaluc\'{i}a}
\newacro{IAC}{Instituto de Astrof\'{i}sica de Canarias}
\newacro{IAS}{Institut d'Astrophysique Spatiale}
\newacro{IAU}{International Astronomical Union}
\newacro{IBIS}{Interferometric Bi-dimensional Spectrometer}
\newacro{IDL}{Interactive Data Language}
\newacro{IMaX}{Imaging Magnetograph eXperiment}
\newacro{INAF}{Istituto Nazionale di Astrofisica}
\newacro{IB}{IRIS bomb}
\newacro{IR}{infrared}
\newacro{IRIS}{Interface Region Imaging Spectrograph}
\newacro{ISAS}{Institute of Space and Astronautical Science}
\newacro{ISP}{Institute for Solar Physics}
\newacro{ISS}{International Space Station}
\newacro{ISSI}{International Space Science Institute}
\newacro{ITA}{Institute for Theoretical Astrophysics}
\newacro{JAXA}{Japan Aerospace Exploration Agency}
\newacro{JSOC}{Joint Science Operations Center}
\newacro{KIS}{Kiepenheuer--Institut f\"{u}r Sonnenphysik}
\newacro{KPNO}{Kitt Peak National Observatory}
\newacro{LASP}{Laboratory for Atmospheric and Space Physics}
\newacro{LC}{liquid cristal}
\newacro{LMSAL}{Lockheed Martin Solar and Astrophysics Labratory}
\newacro{LOS}{line of sight}
\newacro{LTE}{local thermodynamic equilibrium}
\newacro{MC}{magnetic concentration}
\newacro{MCAO}{multi-conjugate adaptive optics} 
\newacro{MDI}{Michelson Doppler Imager}
\newacro{ME}{Milne-Eddington} 
\newacro{MHD}{magnetohydrodynamics}
\newacro{MOMFBD}{Multi-Object Multi-Frame Blind Deconvolution}
\newacro{MPE}{Max--Planck--Institut f\"ur extraterrestrische Physik}
\newacro{MPG}{Max--Planck--Gesellschaft}
\newacro{MPS}{Max Planck Institute for Solar System Research}
\newacro{MSSL}{Mullard Space Science Laboratory}
\newacro{MTF}{modulation transfer function}
\newacro{NAOJ}{National Astronomical Observatory of Japan}
\newacro{NASA}{National Aeronautics and Space Administration}
\newacro{NIST}{National Institute of Standards and Technology}
\newacro{NLTE}{non-local thermodynamic equilibrium}
\newacro{NLFFF}{non-linear force-free field}
\newacro{NOAA}{National Oceanic and Atmospheric Administration}
\newacro{non-E}{non-equilibrium}
\newacro{NSO}{National Solar Observatory}
\newacro{NWO}{Netherlands Organisation for Scientific Research}
\newacro{PHE}{propagating heating event}
\newacro{PRD}{partial redistribution}
\newacro{PROBA2}{PRoject for OnBoard Autonomy}
\newacro{PSBE}{post Saha-Boltzmann extinction}
\newacro{PSF}{point spread function}
\newacro{QS}{quiet Sun}
\newacro{QSEB}{quiet-Sun Ellerman-like brightening} 
\newacro{RAL}{Rutherford Appleton Laboratory}
\newacro{RBE}{rapid blue-shifted excursion}
\newacro{R-MHD}{radiation hydrodynamics}
\newacro{rms}{root mean square}
\newacro{RMS}{root mean square}
\newacro{ROB}{Royal Observatory of Belgium}
\newacro{ROI}{region of interest}
\newacro{RRE}{rapid red-shifted excursion}
\newacro{RTE}{radiative transfer equation}
\newacro{RTSA}{Radiative Transfer in Stellar Atmospheres}
\newacro{SCF}{slender \CaIIH\ fibril}
\newacro{SE}{statistical equilibrium}
\newacro{SB}{Saha Boltzmann}
\newacro{SDO}{Solar Dynamics Observatory}
\newacro{SJI}{slit-jaw image}
\newacro{SLI}{slit image}
\newacro{SNR}{signal-to-noise ratio}
\newacro{SO}{Solar Orbiter}
\newacro{SoHO}{Solar and Heliospheric Observatory}
\newacro{SP}{Spectropolarimeter}
\newacro{SST}{Swedish 1-m Solar Telescope}
\newacro{SUMER}{Solar Ultraviolet Measurements of Emitted Radiation}
\newacro{SUFI}{Sunrise Filter Imager}
\newacro{SVD}{singular value decomposition}
\newacro{SVST}{Swedish Vacuum Solar Telescope}
\newacro{STX}{Solar Telescope X}
\newacro{THEMIS}{T\'{e}lescope H\'{e}liographique pour l'Etude du 
   Magn\'{e}tisme et des Instabilit\'{e} Solaires}     
\newacro{TR}{transition region}
\newacro{TRACE}{Transition Region and Coronal Explorer}
\newacro{TSI}{total solar irradiance}
\newacro{UT}{Universal Time}
\newacro{UV}{ultraviolet}
\newacro{VAULT}{Very high angular resolution ultraviolet telescope}
\newacro{VIRGO}{Variability of solar IRradiance and Gravity Oscillations}
\newacro{VTT}{Vacuum Tower Telescope}    
\newacro{XRT}{X-Ray Telescope}

\long\def\startignore #1\stopignore{}   

\def\rmit#1{{\it #1}}              
           
\def\eg{\rmit{e.g.,}}              
\def\cf{cf.}                       

\def\specchar#1{\uppercase{#1}}    

\def\Halpha{\mbox{H\hspace{0.1ex}$\alpha$}} 

\def\CaIIH{\mbox{Ca\,\specchar{ii}\,\,H}}

\def\level #1 #2#3#4{$#1 \; ^{#2} \mbox{#3} ^{#4}$}   

\def\rmit#1{{\it #1}}         
\def\eg{\rmit{e.g.}}          
\bibpunct{(}{)}{;}{a}{,}{,}   
\def\specchar#1{{\sc{#1}}}    

\begin{document}
\begin{article}
\begin{opening}

\title{In Memoriam Cornelis de Jager}

\author[addressref={aff1,aff2,aff3},corref,email={R.J.Rutten@uu.nl}]
{\inits{R.J.~}\fnm{Robert~J.~ }\lnm{Rutten\textsuperscript{123}}} 
\author[addressref={aff2,aff3}]{\inits{O.}\fnm{Oddbj{\o}rn~}\lnm{Engvold\textsuperscript{23}}}
\author[addressref={aff4}]{\inits{A.C.T.~}\fnm{Adrianus~C.T.~}
\lnm{Nieuwenhuizen\textsuperscript{4}}}

\runningauthor{R.J. Rutten et al.}
\runningtitle{In Memoriam C. de Jager}


\begin{abstract}  
Cornelis (``Kees'') de Jager, co-founder of {\em Solar Physics\/},
passed away in 2021.  
He was an exemplary human being, a great scientist, and had large
impact on our field. 
In this tribute we first briefly summarize his life and career and
then describe some of his solar activities, from his PhD thesis on the
hydrogen lines in 1952 to the book on cycle-climate relations
completed
in 2020.
\end{abstract}
\end{opening}

\begin{figure}
 \centering
  \includegraphics[height=57mm]{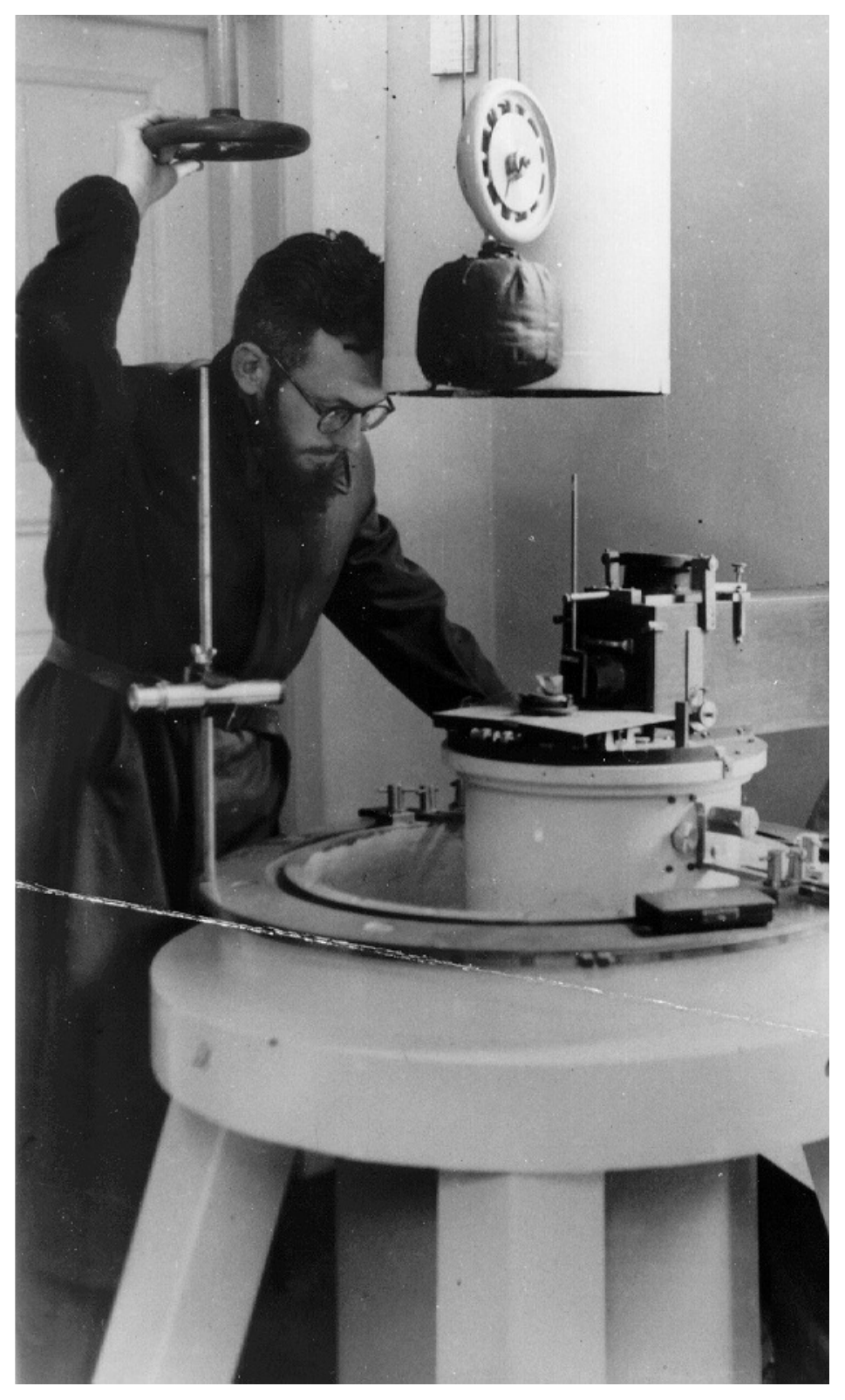}
  \includegraphics[height=57mm]{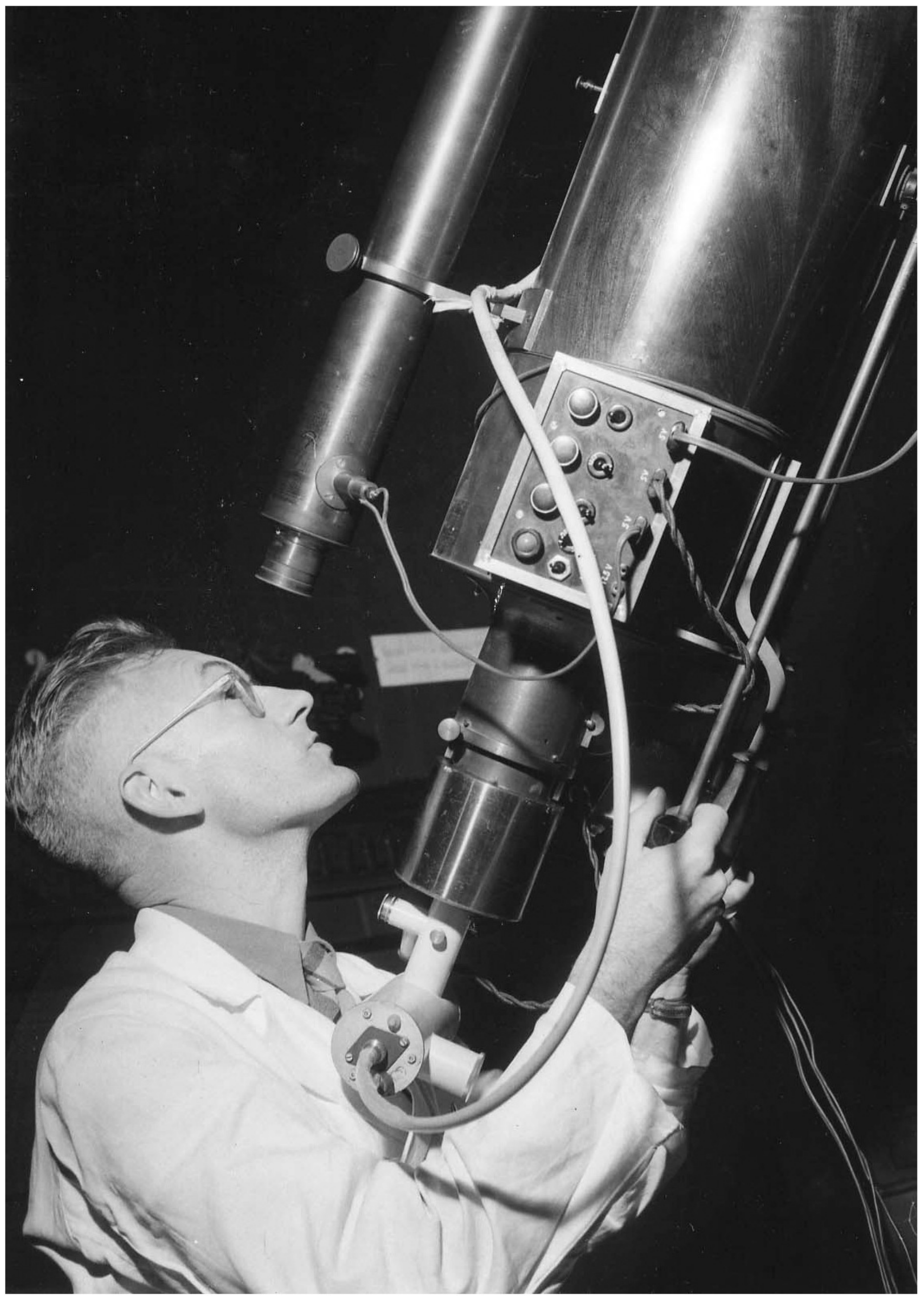}
  \includegraphics[height=57mm]{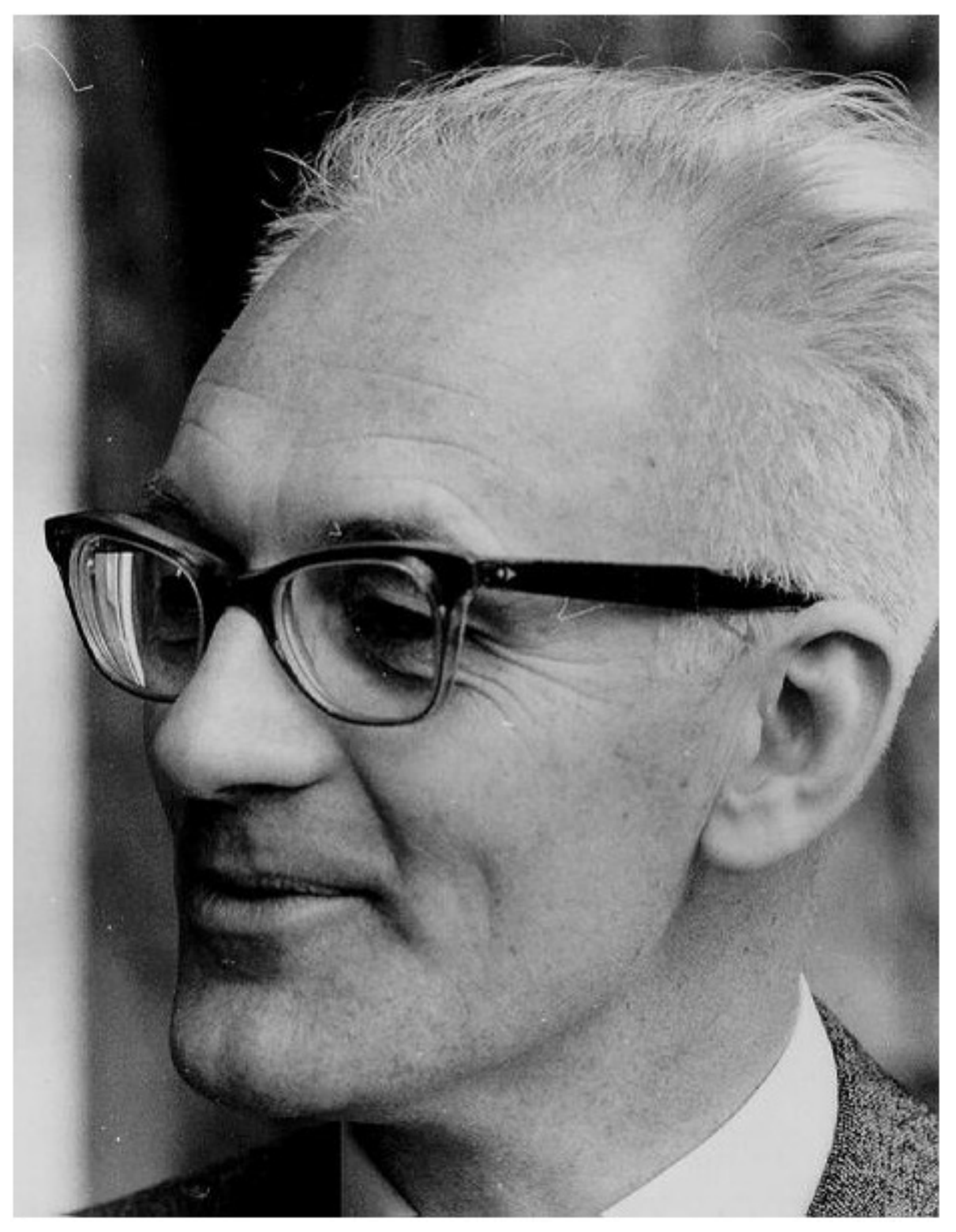}
  \caption[]{{\em Left:\/} Kees at the entrance and exit plane of the
  vertical solar spectrograph at Sonnenborgh in 1944. 
  He adjusts the tilt of the declination mirror of the rooftop
  heliostat feeding the vertical 15-cm refractor spanning two floors;
  the sandbag preloads the clock-controlled hour-angle mirror drive.
  The monk robe is against the cold.
  {\em Center:\/} Kees at the 25-cm Merz refractor at Sonnenborgh in
  1953.
  He used it in his first publication finding fine structure on
  Jupiter's moons (\citeads{1946BAN....10...81D}) and at this time
  built a self-recording photometer for it
  (\citeads{1953BAN....12...93D}) starting a variable star program
  (PhD theses of A.C. de Landtsheer and R.H. van Gent) later utilizing
  a 40-cm telescope built in the Utrecht workshop under Kees' guidance
  (Minnaert required solid training in using workshop machinery) and
  stationed in Greece and Switzerland. 
  In the 1950s scientists wore white lab coats. 
  Kees' hair turned white prematurely, perhaps contributing to
  diplomatic feats as Taiwan + China ICSU membership.
  {\em Right:\/} Kees at his Bilderberg meeting in 1967 (photo
  R.J.\,Rutten).}
\end{figure}

\paragraph*{Life and career.}
De Jager was born on April 29, 1921 in Den Burg on the island of Texel
in The Netherlands and passed away in the same village on May 27,
2021. 
He was a giant in our field and far beyond, ranging widely through the
spectrum in solar physics and through stellar physics, educating many
astrophysicists, traveling widely as an inventive and effective
science diplomat, founding institutes, journals and societies,
intensely active in popularization -- and running marathons as well. 
Always friendly, cooperative, attentive, positive and optimistic. 
A wonderful person and a great scientist.

At the occasion of his 75th birthday {\em Solar Physics\/} published a
special volume in which he wrote a highly recommended,
characteristically well-written and humorous overview of his rich life
and career until then ({\em ``A red and a white star''\/}
\citeads{1996SoPh..169..443D})\footnote{In addition he wrote monthly
reminiscence columns in the Dutch amateur journal {\em Zenit\/} (which
he co-founded when reorganizing Dutch amateur astronomy) from 2006 and
collected 180 in 2014 and 2020 in two books {\em Terugblik\/} (looking
back, De Jager \citeyear{deJager-Terugblik1},
\citeyear{deJager-Terugblik2})
\href{https://zenitonline.nl/product-categorie/boeken}{[order page]}.
They are a marvelous read -- in Dutch.}.
We refer to this for detail on his happy childhood in Indonesia,
studentship at Utrecht University during World War (hiding from
deportation to Germany at Sterrewacht Sonnenborgh), graduation in
1946, PhD thesis with M.G.J.~Minnaert in 1952, professorship in 1960,
succeeding Minnaert as observatory director in 1963 and moving into
the director's mansion at Sonnenborgh with his wife Duotsje (always
called Doetie) and their four children, and also for detail on his
research and on his founding astrophysics at the Brussels Free
University, directorship of the Utrecht Observatory with rapid
expansion from 5 to 50 employees, starting the Utrecht space research
laboratory quickly growing to over 100 employees, involvement in ESRO,
ESA, the Dutch astronomical satellite ANS and more space
instrumentation, starting {\em Space Science Reviews\/} and
co-starting {\em Solar Physics\/} (more below), co-founding JOSO (more
below), being IAU General Secretary, president of COSPAR twice,
president of ICSU,
and much more. 
His role in ESRO and early ESA is well described by
\citetads{1996SoPh..169..233B} in the same {\em Solar Physics\/} issue.

After turning 75 he turned Sterrenwacht Sonnenborgh into a successful
science museum, during many years giving monthly lectures on anything
astronomical in the news during the weeks before and turning these
into books, also after his return to Texel in 2003. 
He also co-founded and chaired the Dutch and European skeptical
societies debunking pseudoscience\footnote{Entry 
\href{https://en.wikipedia.org/wiki/Kees_de_Jager}{Kees de Jager} of
Wikipedia (English) serves his tongue-in-cheek descriptions of Cyclosophy,
the bicycle-based holistic four-dimensional religion that he created
for the New Age of Aquarius. 
Earlier he showed that many dimensions of his Sonnenborgh home had
cosmic proportions well beyond the Cheops pyramid that forecast the
end of the world on March 16, 2674 at 03:56\,UT.}.

On Texel he delved into local history, also lecturing on that, and
became guest researcher at the Dutch institute for maritime research (NIOZ)
there, turning to relations between solar activity and climate in a
dozen publications
while also co-authoring further studies on hypergiants. 
He put the first together in {\em Solar magnetic variability and
climate\/} in 2020 (more below).

Among many other awards he received the Karl Schwarzschild Medal in
1974, the Prix Jules Janssen and the Gagarin Medal in 1984, the
Ziolkowskiy Medal in 1987, the Hale Prize and the RAS Gold Medal in
1988, honorary citizenship of Texel in 2006 -- and he also held the age
record for Dutch marathon runners (New York at 75 after starting
long-distance running at 50).

At Utrecht University he was PhD (co-)adviser to M. Kuperus (1965),
L.D. de Feiter (1966), E.P.J. van den Heuvel (1968), J. Roosen (1968),
A,C. Brinkman (1972), R.C.P. van Helden (1972), H.F. van Beek (1973),
H.J.G.L.M. Lamers (1974), P. Hoyng (1975), G.C.M. Reijnen (1976),
K.A. van der Hucht (1978), E.H.B.M. Gronenschild (1979), A.J.F. den
Boggende (1979), J. Heise (1982), A. Duyveman (1983), A.C. de
Landtsheer (1983), P.P.L. Hick (1988) and R.H. van Gent (1989). 
At the Free University of Brussels to C.W.H. De Loore (1968),
E.L.J. van Dessel (1973), M. Burger (1976) and A. Lobel (1997).

\paragraph*{PhD thesis.}
Kees\footnote{We use the familiar ``Kees'' on purpose.
Everybody at the Utrecht observatory and Kees' space research laboratory
called him that (in the 1960s the former called the latter
``ruimtekezen'' = space Keeses), in contrast to formal
M.G.J.~Minnaert (1893\,--\,1970) who was called ``Professor'' by
everybody including professor Kees, 
even by J.~Houtgast (1908\,--\,1982) whom Minnaert
called ``Mijnheer'' throughout their four-decade collaborations.} 
used his years of hiding at Sterrewacht Sonnenborgh to work through
most of the astronomy--physics--math curriculum and graduated in 1946. 
Minnaert suggested a PhD thesis on the solar Balmer lines using the
spectrograph at Sonnenborgh (that Minnaert had moved from the Physics
Laboratory when reviving the observatory in the 1930s).
Kees had hoped to venture further out in the universe\footnote{Kees
coined the beautiful ``oerknal'' for ``big bang'' in the Dutch
language.} then just eight minutes but set himself to the task.

The resulting thesis {\em The hydrogen spectrum of the sun\/} is now
available at ADS
(\citeads{1952RAOU....1.....D}). 
The description of his data gathering is a good reminder of practices
before electronic detection and computing.
At Utrecht he collected center-to-limb spectra of many Balmer lines on
a hundred photographic plates. 
After development in the observatory darkroom followed scanning with
Houtgast's ingenious double-galvanometer intensity-recording
microdensitometer (diagram in the introduction to the Utrecht Atlas by
\citeads{1940pass.book.....M}) 
after determination of non-linear plate density to intensity
calibration curves from spectra taken through a step weakener, and
then applying extensive corrections for grating ghosts, straylight and
the instrumental profile. 
The final results are line profiles given in tables. 
He similarly collected \Halpha\ spectroheliograms and off-limb spectra
at Meudon and Paschen and Brackett line profiles at Jungfraujoch.
Then follows a careful comparison with other published Balmer-line
profiles resulting in summarizing profile figures drawn by the
observatory drafter.
Kees then constructed a sequence of solar-atmosphere models starting
by fitting visible continua and their limb darkening, then
hydrogen-line wings for deep layers and finally the line cores for
high layers. 
All this per mechanical calculator, slide rule, logarithm tables,
(logarithmic) graph paper, mental arithmetic (also in logarithms!).
The main problem was that he could not reconcile the wings with the
cores without adding ad-hoc line broadening increasing with wavelength
-- too much for invoking deficiencies of Stark broadening theory.
The last chapters address chromospheric modeling with the off-limb
profiles of \Halpha, admittedly more speculative and suggesting that
spicule motions must be an important but unknown ingredient.

In hindsight Kees commented that it was all wrong by assuming LTE. 
Indeed the hydrogen lines are scattering lines with attendant
source-function darkening and of course chromospheric inhomogeneity
is a major ingredient. 
Nowadays the approach would be to compare high-resolution observations
and simulations, with 3D spectral synthesis.
Such studies have not yet been done for multiple Balmer, Paschen and
Brackett lines as comprehensively as Kees did in his pioneering
thesis.

\paragraph*{Solar atmosphere.}
Kees continued working on the Paschen and Brackett lines with L.~Neven
at the Royal Observatory of Belgium\footnote{ADS lists 35 joint
publications spanning 1950\,--\,1982 (46 in total for Neven). 
During 1961\,--\,1972 Kees taught astronomy weekly at what became the
Vrije (non-catholic flemish) Universiteit Brussel, going there (by
train, Kees had no driving license) on Fridays to teach his courses
({\em Stellar structure and evolution\/} the most inspiring), spending
the night at the observatory which had guest rooms, and on Saturdays
working there with Neven before his return.} 
and refining solar atmosphere models with H.~Hubenet and
J.R.W.~Heintze at Utrecht.  
The highlight became his ``International study week'' in 1967 at the
famous hotel De Bilderberg in which he gathered foremost solar
physicists and promising youngsters
to jointly derive a definitive model.
The decision there was to use only the continuum, lines being too
difficult, and there was much debate about whether microturbulence,
the problematic adjustment parameter to broaden lines as much as
observed, increases or decreases with height. 
The result was the Bilderberg Continuum Atmosphere of
\citetads{1968SoPh....3....5G}\footnote{For 
more on the Bilderberg meeting and these modeling efforts see
\citetads{2002JAD.....8....8R} 
in a JAD issue celebrating Kees' 80th birthday.}.

Kees then quit these efforts. 
One-dimensional continuum modeling moved to Harvard, first the HSRA of
\citetads{1971SoPh...18..347G} 
and then the long sequence of models constructed by E.H.~Avrett and
coworkers, earning the ``standard-model'' standing that Kees had
anticipated as major need.
At Utrecht solar atmosphere studies were taken over in the 1970s by
C.~Zwaan (1928\,--\,1999) and students (adding solar--stellar activity),
in the 1990s by R.J.~Rutten and students, in the 2000s by C.U.~Keller
and students.

Meanwhile Kees maintained his interest in turbulence, proposing better
descriptions (mesoturbulence next to micro and macro) and also in
stellar applications when he extended his research beyond solar
physics to mass loss, supergiants and
hypergiants\footnote{Hypergiants were named after a Dutch cartoon
character proposed by A.M.~van Genderen.
Kees' last entry on ADS is \citetads{2019A&A...631A..48V}.
The co-authors include H.~Nieuwenhuijzen who was Kees' main coworker
after his 1986 ``retirement'' and co-author on 18 ADS entries,
mostly on stellar mass loss and giant stars and more recently on
cycle--climate connections until A.C.T.~Nieuwenhuizen took over.}
prompted by starting stellar ultraviolet spectroscopy as a non-solar
space research activity with the S59 spectrometer on ESRO's TD-1A
satellite (\citeads{1974Ap&SS..26..207D}) 
and writing {\em The brightest stars\/}
(\citeads{1980GAM....19.....D})\footnote{ADS 
lists 166 citations but no full-text source. 
However, Springer sells a scan, as for the 27 other {\em Geophysics
and Astrophysics Monographs\/} published by Reidel.}
to thoroughly master the subject.

\begin{table} 
\caption[]{\label{tab:handbuch}
Overview of  {\em Structure and Dynamics of the Solar Atmosphere\/}.  
These are only the major headers.
Third column: topic numbers (paragraph headers).
Last column: page numbers printed in \citetads{1959HDP....52...80D}.}
\begin{tabular}{lllr}
\hline
Part A & The undisturbed photosphere and chromosphere && 80 \\
       & I.  The undisturbed photosphere & 2-14 & 80 \\
       & II.  The chromosphere & 15-30 & 106 \\
Part B & The disturbed parts of the photosphere and chromosphere && 151 \\ 
       & I. Sunspots & 31-39 & 151 \\
       & II. Photospheric and chromospheric faculae & 40-47 & 173 \\
       & III. Flares and associated phenomena & 48-59 & 191 \\
       & IV. Filaments and prominences & 60-69 & 224 \\
Part C & The corona && 248 \\
       & I. Optical observations & 70-81 & 248 \\
       & II. Radio emission from the Sun & 82-95 & 283 \\
Part D & Solar rotation and the solar cycle & 96-105 & 322 \\
Part E & Solar and terrestrial relationships & 106-110 & 344 \\
\hline
\end{tabular}
\end{table}

Kees' trick of writing a comprehensive review to learn a field
inside-out started with {\em Structure and Dynamics of the Solar
Atmosphere\/}, a 280-page overview establishing him as major solar
physicist (\citeads{1959HDP....52...80D}).\footnote{Springer sells
a scan but without activating crossreferences nor 
hyperlinking citations.}
It furnishes a brilliant, complete inventory of solar physics in Kees'
clear style at the time when the computer and space eras were
starting. 
Over 100 succinct essays surveying the state of solar physics in a
wide range of topics.
Table~1 
summarizes the contents.
Compiling it gave Kees detailed knowledge of the rich complexities
that the solar atmosphere has on offer.\footnote{When Thomas and Athay
wrote {\em Physics of the Solar Chromosphere\/} mostly debating
NLTE line formation
(\citeads{1961psc..book.....A}, 
wrong author order at ADS) 
Kees wrote a scathing review (at ADS wrongly labeled
\citeads{1962ZA.....55...66T} with the tail wrongly in
\citeads{1962ZA.....55...70W}) that a glance through an \Halpha\
filter makes one wonder about waves, shocks and magnetism --
altogether ignored in the book.
Indeed the book and Kees' review describe utterly different stars. 
The five-page critique also demonstrates Kees' fluency in German --
the high school he attended with Doetie in Surabaya held the high
standard of Dutch education at the time. 
He also learned proper Dutch there; his principal languages had been
the Texel dialect and Minahasan.}  

\paragraph*{Flares.}
Flare\footnote{The Dutch term ``zonnevlam'' came also from Kees.}
monitoring started in The Netherlands in the 1950s with Lyot-filter
\Halpha\ observations by the Dutch telephone service for predicting
solar disturbances of overseas radio connections.
Kees organized a direct telephone line to get warned the moment an
\Halpha\ flare was detected and streamlined the Sonnenborgh start-up
procedure to yield \Halpha\ spectrograms within a minute. 

At this time Kees also became interested in solar radio ``storms''
during flares and measured them on fast recordings collected by
C.L.~Seeger at Cornell
(\citeads{1958VKAWA..21....1D}). 
This led to the Utrecht solar radio group that blossomed in the 1960s
and 1970s\footnote{PhD theses T.~de Groot (1966), J.~Roosen (1968),
J.~van Nieuwkoop (1971), J.~Rosenberg (1973), J.M.E.~Kuijpers (1975),
A.~Kattenberg (1981) and C.~Slottje (1982).}, culminating in J. van
Nieuwkoop's fast 60-channel radio spectrograph at the Dwingeloo 25-m
antenna (\eg\ millisecond flare spikes in
\citeads{1978Natur.275..520S}). 

At the launch of {\em Sputnik\/} Kees recognized the potentialities
of space research, became involved in starting ESRO (later ESA) and
convinced Dutch funding agencies to start the Utrecht space research
laboratory (ROU, later national SRON) with X-ray flare studies as
major topic, plus stellar ultraviolet spectroscopy and non-solar X-ray
research.
The ESRO-2 satellite (1968) got a soft X-ray monitor for flare studies
but already before launch Kees realized that soft X-rays show only the
gradual phase after ignition, while he wanted to study triggers in the
onset phase always missed in his \Halpha\ spectrometry (which he
therefore never published).  
The ESRO TD-1A satellite (1972) got a hard X-ray monitor but there was
a clear need for imaging to locate impulsive-phase bursts. 
This was realized with a train of collimator grids in the {\em Hard
X-ray Imaging Spectrometer\/} (HXIS,
\citeads{1980SoPh...65...39V}) 
on the {\em Solar Maximum Mission\/} (SMM), which was launched early in
1980. 
The finding of association between \Halpha\ flare kernels and
impulsive X-ray bursts by
\citetads{1981ApJ...246L.155H} 
for the May 21, 1980 flare remains Kees' most cited (presently 223) ADS
entry. 
The May 30, 1980 limb flare became known as ``Queens flare'' because
that day Dutch Queen Juliana was succeeded by Queen Beatrix
(\citeads{1981ApJ...244L.157V}, 
\citeads{1983SoPh...84..205D}). 
Later that year SMM suffered fuse failure and was put in standby until
partial repair by shuttle astronauts in 1984.
HXIS was not revived but the flares observed during 1980 helped
establish the standard flare scheme of loop-top reconnection with
particle-beam footpoint heating (\eg\
\citeads{1989SoPh..122..263M}). 
Kees then got involved in {\em Yohkoh\/} flare studies (\eg\
\citeads{1996SSRv...77....1S}). 
The collimator mask technology (nowadays ``photon sieves'') that was started
with HXIS and precursor projects trying Fresnel zone plates became a
major SRON asset, also in X-ray spectroscopy.
 
\paragraph*{Solar Physics}
The Sun's central role in stellar science motivated Kees to start the
journal {\em Solar Physics\/} in 1967. 
A.~Reidel, the publisher with whom Kees had started {\em Space Science
Reviews\/} (\citeads{1962SSRv....1....5D} with a wise Erasmus opener)
and the {\em Astrophysics and Space Science Library\/} book series
with the proceedings {\em The Solar Spectrum\/} of the 1963
Minnaert-farewell symposium\footnote{For 
photographs taken by H.~Nieuwenhuijzen see his
\href{https://robrutten.nl/nieuwenhuijzenshots/sym1963/album.html}
{symposium album}.}
(\citeads{1965ASSL....1.....D}), urged Kees that another journal from
him would be welcome -- but Kees hesitated until Zden{\v{e}}k
\v{S}vestka reacted enthusiastically and proposed teaming up. 
They then started the journal together (editorial
\citeads{1967SoPh....1....3D}). 
They became a strong editorial team, further strengthened when Bob
Howard joined them in 1987. 
Kees served as co-editor during 29 years and then, as honorary member
of the Editorial Board for the next 25 years, kept providing essential
contributions and support to the journal.
It became a mainstay in solar physics.

\paragraph*{JOSO.}
The need for higher-resolution observations of the Sun became a major
concern in the 1960s. 
At the 1967 IAU Symposium {\em Structure and Development of Solar
Active Regions\/} in Budapest organized by K.O.~Kiepenheuer, he and Kees
agreed to form a collaboration to find an outstanding site for a new
European solar observatory. 
In 1969 a group of leading European solar physicists from seven
countries established a convention to collaborate towards this goal
under the name {\em Joint Organization for Solar Observations\/} (JOSO). 
All expenses were covered by voluntary contributions from
adhering institutes. 
Kiepenheuer, Kees and Edith M\"uller were successive presidents. 
Labor-intensive site-testing campaigns became possible with
enthusiastic participation of solar scientists and students from all
over Europe.
After testing close to 40 sites in the Mediterranean and Atlantic
coastal areas, the searches focused on mountain sites in the Canary
Islands with Iza\~{n}a in Tenerife and Roque de los Muchachos on La
Palma the most outstanding. 

Subsequently JOSO became involved in developing new telescope
technology for these excellent sites. 
Kees' Utrecht colleagues C.~Zwaan and R.H.~Hammerschlag tested the
non-vacuum open-telescope principle by installing the 45\,cm {\em
Dutch Open Telescope\/} (DOT) on La Palma. 
The later German 1.5\,m GREGOR at Iza\~{n}a and the 4.2\,m DKIST on
Maui, Hawaii are based on open solutions, also chosen for the 4\,m
{\em European Solar Telescope\/} (EST) currently in development for
the DOT site on Roque de los Muchachos. 

JOSO has also been actively involved in coordinating observations
between ground-based solar facilities and instruments in space as
SOHO and TRACE. 
Overall, the somewhat unusually organized 
JOSO has contributed notably to solar physics in past decades.

\paragraph*{Cycle and climate.}
As a guest worker at the Dutch institute for maritime research NIOZ on
Texel, Kees returned to solar physics, at their suggestion 
studying possible relations between solar activity and Earth
climate. 
In \citetads{1981PRNAA..84..457D} 
he had already plotted Dutch winter temperatures (derived from
ice-skating\footnote{Growing up in Indonesia had made Kees a
non-skater whereas J.H.\,Oort took the Leiden observatory on long
skating tours at every opportunity.} 
records) against sunspot numbers and found at most weak correlation,
concluding ``More thorough efforts, using sophisticated statistical
methods, will not yield more reliable results''.

Characteristically he started with a thorough review
(\citeads{2005SSRv..120..197D}, 
44 pages, 44 citers,
abstract concluding ``The future of such a chaotic system is
intrinsically unpredictable'') while skeptically
repudiating planetary attraction periodicities as significant
contribution
(\citeads{2005SoPh..229..175D}; 
\cf\
\citeads{2013JASTP.102..372C}). 
Then \citetads{2006JASTP..68.2053D} 
concluded that ultraviolet irradiance variations have larger effect
than cosmic ray shielding variations. 
The importance of energetic particle and spectral irradiance variations
as potentially more climate-effective than the minute total irradiance
modulation is now recognized by the IPCC.


 \begin{figure}
 \centering
  \includegraphics[height=64mm]{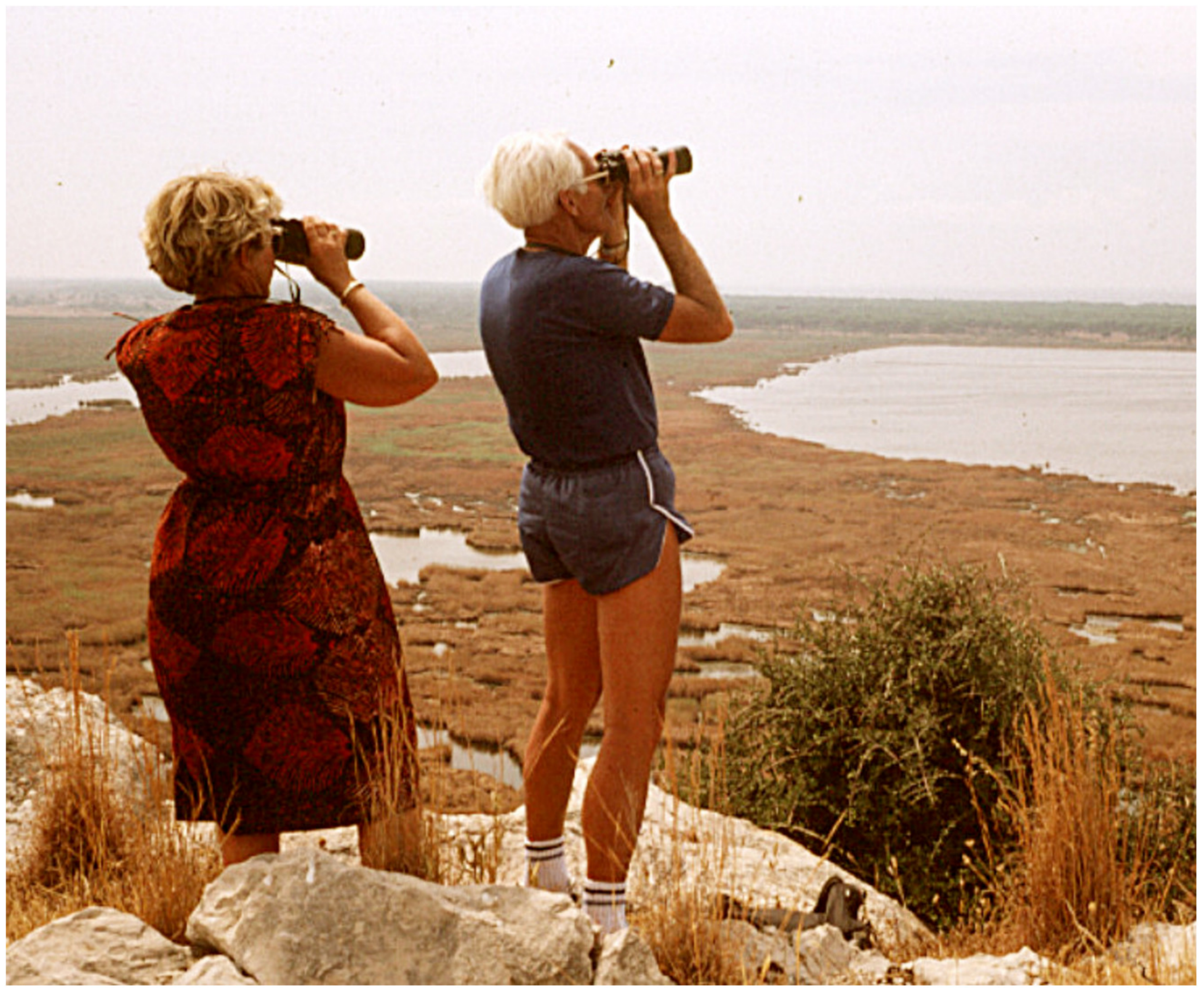}
  \includegraphics[height=64mm]{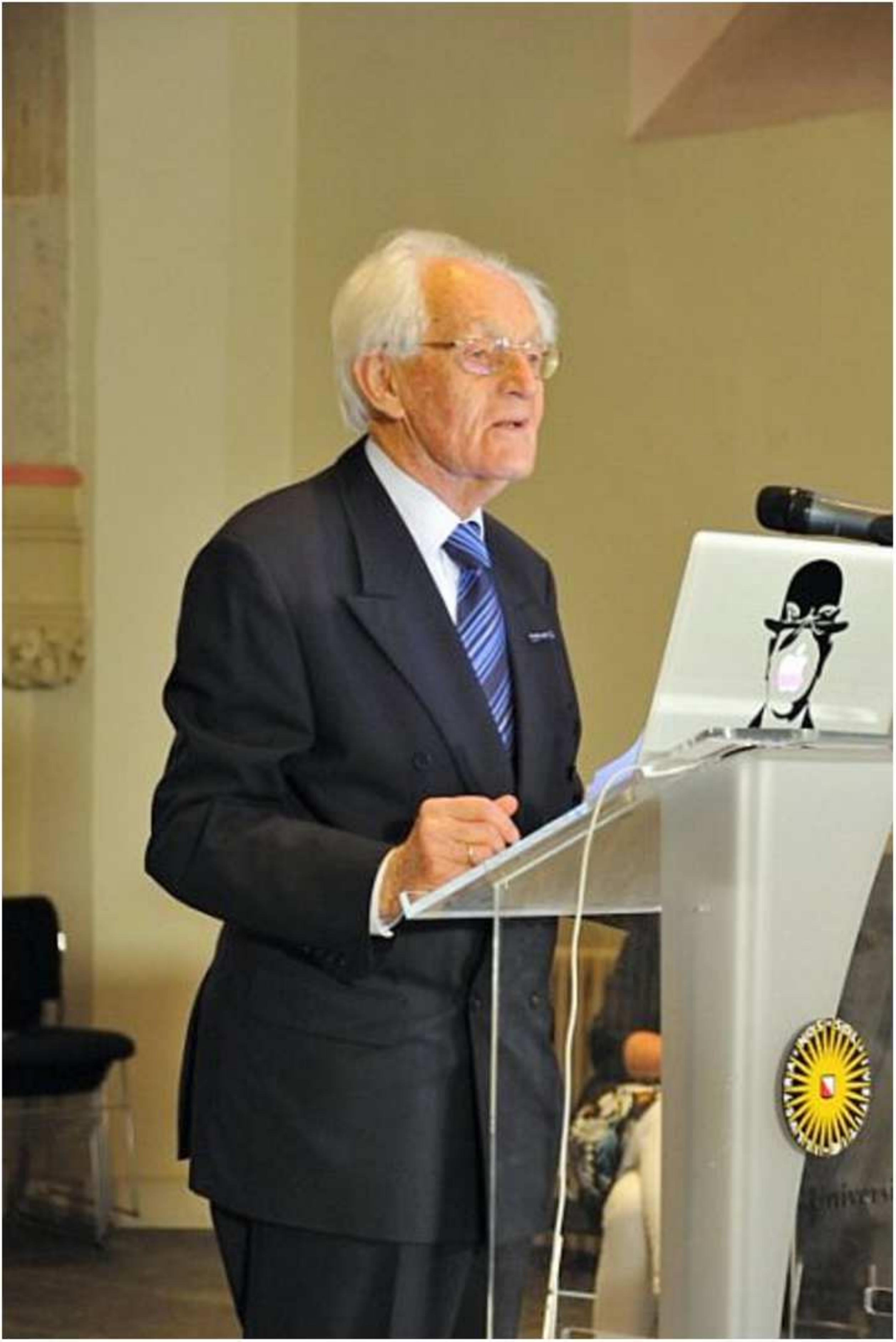}
  \caption[]{%
  {\em Left:\/} Doetie and Kees watching terns at Propokou Lagoon on
  the coast west of Patras during the 1982 General Assembly of the
  IAU.  
  They were avid birders (Texel offers the best opportunities in The
  Netherlands) and spent all free time on their travels on such
  excursions. 
  Here they arrived on mopeds, new to them; Kees in running attire
  (photo R.J.~Rutten).
  Earlier that week Kees gave an Invited Discourse on solar flares in
  the Roman Odeon in Patras, a spectacular setting.
  {\em Right:\/} Kees on April 4, 2012 at the farewell ceremony in the
  medieval Utrecht university hall (where the Treaty of Utrecht
  started The Netherlands in 1579; Utrecht astronomy started in 1642,
  see \citeads{2013ASPC..470...15D}). 
  Kees gave an emotional and moving speech: ``Here in Utrecht I raised
  a beautiful daughter. 
  I enjoyed watching her develop and grow into splendid maturity --
  but then she was brutally murdered''.
  The Utrecht University logo has the Sun as emblem (yellow {\em Sol
  Iustitiae Illustra Nos\/} at lower right) but now undeservedly
  (photo H.~Nieuwenhuijzen in his
  \href{https://robrutten.nl/nieuwenhuijzenshots/2012-SIU-afscheid/album.html}
  {farewell album}).}
\end{figure}


Kees then started a collaboration with S.~Duhau (Buenos Aires)
inspired by her inclusion of polar fields and her cycle phase diagram
plotting wavelet-fitted geomagnetic versus sunspot number modulations
in \citetads{2002GeoRL..29.1628D}. 
It lasted the rest of his life and resulted in ten more publications
recapitulated in {\em Solar magnetic variability and
climate\/} 
(\citealp{2020dejagervariability}).\footnote{ 
Submitted to Springer in June 2020 but in autumn Kees became
impatient at lack of response and sent it instead to the
publisher of {\em Zenit\/} and his {\em Terugblik\/} books
\href{https://robrutten.nl/rrweb/rjr-pubstuff/bookcdej/solar-variability-and-climate-content.pdf}{[contents overview]}
\href{https://zenitonline.nl/product-categorie/boeken}{[order page]}.}  
Main themes are:
\begin{itemize} \vspace{-1ex}  
\item[{\bf --}] cycle phase diagrams plotting modulation of toroidal
magnetism (spots, plage, active network) with sunspot number maxima as
proxy against modulation of poloidal magnetism (polar mixed-network
faculae, polar coronal bright points) with geomagnetic minima as
proxy. 
These show looped tracks with time that differ between grand minima
(as the Maunder minimum), grand maxima (as the recent one) and regular
episodes, with the transitions related to the Gleissberg cycle;

\item[{\bf --}] wavelet decomposition identifying eight oscillatory
cycle modes (including planets after all) permits extrapolation to
predict that a regular episode started around 2006 with maxima
specified up to 2140;


\item[{\bf --}] regression analysis correlating the toroidal and
poloidal solar proxies with northern-hemisphere Earth temperatures
using 18-year smoothing yields good correspondence with 
0.3~degree amplitude until the onset of anthropogenic heating around 1920.
Up to then the toroidal component contributed about 43\,\%, the poloidal
component about 32\,\%. 
The remainder is tentatively attributed to a non-cyclic surface
dynamo producing small magnetic concentrations coalescing into
mixed-polarity network. 
The best match is obtained using a 1\,--\,2 decade delay that recently
diminished, perhaps related to glacier area.

\end{itemize} \vspace{-0.5ex}%
Kees liked to say that he should get another century to see how the
cycle and these predictions fare -- failure would be welcomed as
reason to delve deeper. 
His work is already motivation to delve deeper.


\paragraph*{Utrecht farewell.}
An unexpected drama unfolded abruptly in 2011.
For unknown reasons the board of Utrecht University 
suddenly decided to close the prospering Sterrekundig Instituut
Utrecht (renamed from ``Sterrewacht'' after leaving Sonnenborgh in
1987).
Director C.U.~Keller orchestrated a deal in which most Utrecht
astronomers were relocated to astronomy departments at Leiden,
Nij\-megen and Amsterdam. 
Kees' space research laboratory SRON sought better connections, first
in Amsterdam, then moved to the Leiden campus.
The DOT is mothballed since.
Utrecht solar physics, effectively Dutch solar physics, is gone.
However, a university board terminating its astronomy program
cannot undo the sky-high record of a most illustrious scientist built
in his hundred rounds around the Sun.

\bibliographystyle{spr-mp-sola} 
\bibliography{springerrepair,rjrfiles,adsfiles} 

\end{article} 
\end{document}